\documentclass[amssymb,prl,twocolumn,floats]{revtex4}
\usepackage{bm}
\usepackage{graphicx}
\begin{document}

\title{Ferromagnetic semiconductors}
\author{Tomasz Dietl}
\affiliation{Institute of Physics, Polish Academy of Sciences, \\
al. Lotnik\'{o}w 32/46, PL-02-668 Warszawa, Poland}

\begin{abstract}
The current status and prospects of research on ferromagnetism in
semiconductors are reviewed. The question of the origin of
ferromagnetism in europium chalcogenides, chromium spinels and,
particularly, in diluted magnetic semiconductors is addressed. The
nature of electronic states derived from 3d of magnetic impurities
is discussed in some details. Results of a quantitative comparison
between experimental and theoretical results, notably for Mn-based
III-V and II-VI compounds, are presented. This comparison
demonstrates that the current theory of the exchange interactions
mediated by holes in the valence band describes correctly the
values of Curie temperatures $T_{\mbox {\small C}}$, magnetic
anisotropy, domain structure, and magnetic circular dichroism. On
this basis, chemical trends are examined and show to lead to the
prediction of semiconductor systems with $T_{\mbox {\small {C}}}$
that may exceed room temperature, an expectation that are being
confirmed by recent findings. Results for materials containing
magnetic ions other than Mn are also presented emphasizing that the
double exchange involving hoping through d states may operate in
those systems.
\end{abstract}
\maketitle


\section{Introduction}

Because of complementary properties of semiconductor and
ferromagnetic material systems, a growing effort is directed toward
studies of semiconductor-magnetic nanostructures. Applications in
in optical modulators and insulators, in sensors and memories
\cite{deBo02}, as well as for computing using electron spins
\cite{Loss02} can be envisaged. The hybrid nanostructures, in which
both electric and magnetic field are spatially modulated, are
usually fabricated by patterning of a ferromagnetic metal on the
top of a semiconductor \cite{Tana02} or by inserting ferromagnetic
dots or layers into a semiconductor matrix \cite{Akin02}. In such
devices, the stray fields can control charge and spin dynamics in
the semiconductor. At the same time, spin-polarized electrons in
the metal can be injected into or across the semiconductor
\cite{John02}. Furthermore, the ferromagnetic neighbors may affect
semiconductor electronic states by the ferromagnetic proximity
effect even under thermal equilibrium conditions. Particularly
perspective materials in the context of hybrid structures appear to
be those elemental or compound ferromagnets which can be grown in
the same reactor as the semiconductor counterpart.

However, already the early studies of Cr spinels \cite{Crai75},
rock-salt Eu- \cite{Kasu68,Meth68,Wach79,Naga83,Maug86} and
Mn-based \cite{Stor97} chalcogenides led to the observation of a
number of outstanding phenomena associated with the interplay
between ferromagnetic cooperative phenomena and semiconducting
properties. The discovery of ferromagnetism in zinc-blende III-V
\cite{Ohno92,Ohno96a} and II-VI \cite{Haur97,Ferr01} Mn-based
compounds allows one to explore physics of previously not available
combinations of quantum structures and magnetism in semiconductors.
For instance, a possibility of changing the magnetic phase
isothermally, by light \cite{Haur97,Kosh97,Koss00} or by the
electric field \cite{Ohno00,Bouk01}, was put into the evidence in
(In,Mn)As/(Al,Ga)Sb \cite{Kosh97,Ohno00} and (Cd,Mn)Te/(Cd,Zn,Mg)Te
\cite{Haur97,Koss00,Bouk01} heterostructures. The injection of
spin-polarized carriers from (Ga,Mn)As to a (In,Ga)As quantum well
in the absence of an external magnetic field was demonstrated, too
\cite{Ohno99b}. At the same time, outstanding phenomena, known from
the earlier studies of metallic multilayer structures, have also
been observed in ferromagnetic semiconductors, examples being
interlayer coupling \cite{Chib00,Szus01}, exchange bias
\cite{Liu01}, giant \cite{Chib00} and tunneling magnetoresistance
\cite{Tana01}. It is then the important challenge of materials
science to understand the ferromagnetism  in these compounds and to
develop functional semiconductor systems with the Curie
temperatures $T_{\mbox{\small{C}}}$ exceeding comfortably the room
temperature.

We begin this review by describing briefly, in Sec.~2, various
families of ferromagnetic semiconductors and theoretical models
proposed to explain the nature of relevant spin-spin exchange
interactions. We then discuss, in Sec.~3, energies and character of
electronic states derived from 3d shells of magnetic impurities in
II-VI and III-V semiconductors, which provide information on the
charge and spin states as well as on the effect of the magnetic
constituent on the carrier concentration, also in the presence of
co-doping by shallow donors or acceptors. In Sec.~4, we outline the
main ingredients and limitations of the mean-field Zener model put
recently forward to describe quantitatively the hole-mediated
ferromagnetism in tetrahedrally coordinated semiconductors
\cite{Diet97,Diet00,Diet01b}. Results of a quantitative comparison
between experimental and theoretical results for Mn-based III-V and
II-VI compounds are shown in Sec.~5. This comparison demonstrates
that the current theory of the exchange interactions mediated by
holes in the valence band describes correctly the values of
$T_{\mbox {\small C}}$, magnetic anisotropy, and domain structure.
Finally, in Sec.~6, we present the theoretically predicted chemical
trends and discuss various suggestions concerning the design of
high-temperature ferromagnetic semiconductors.

The novel physics of magnetic heterostructures constitutes the
topic of the next paper \cite{MacD02} in this compendium and,
therefore, is not discussed here. There are, of course, a number of
other recent review articles describing many aspects of
ferromagnetic III-V \cite{Ohno99a,Twar00,Mats01b}, II-VI
\cite{Cibe02,Koss02}, and IV-VI materials \cite{Stor97}, which are
not touched upon in this paper.

\section{Families of ferromagnetic semiconductors and relevant
spin-spin exchange interactions}

Manganites (perovskite: (La,Sr)MnO$_3$ and related materials),
which show colossal magnetoresistance, are magnetic semiconductors,
whose studies have been particularly active over the recent years
\cite{Coey99,Toku00}. Their ferromagnetic order, beginning at
~350~K, is brought by the double-exchange interaction involving
on-site Hund's ferromagnetic spin coupling and hopping of d
electrons between Mn$^{3+}$ and Mn$^{4+}$ ions.

The family of magnetic semiconductors encompasses also europium and
chromium chalcogenides (rock-salt type: EuS, EuO and spinels:
CdCr$_2$S$_4$, CdCr$_2$Se$_4$), for which the Curie temperature
$T_{\mbox{\small{C}}}$ does not exceed 100~K. In the case of
rock-salt Eu compounds, there appears to be a competition between
antiferromagnetic cation-anion-cation and ferromagnetic
cation-cation superexchange \cite{Wach79}. The latter can be traced
back to the ferromagnetic s-f coupling, and the presence of s-f
hybridization, which is actually stronger than the p-f
hybridization due to symmetry reasons \cite{Wach79,Diet94c}. In
such a situation, the lowering of the conduction band associated
with the ferromagnetic order enhances the energy gain due to
hybridization. The Cr-spinels represents the case, in which the d
orbitals of the two cations are not coupled  to the same p orbital,
which results in a ferromagnetic superexchange.

In europium and chromium chalcogenides discussed above, the
presence of carriers can affect $T_{\mbox{\small{C}}}$ but is not
necessary for the appearance of the ferromagnetic order. In
contrast, in the case of Mn-based IV-VI \cite{Stor97}, III-V
\cite{Ohno98}, and II-VI \cite{Diet99} diluted magnetic
semiconductors (DMS) the ferromagnetism can be observed provided
that the hole concentration is sufficiently high.  According to
experimental and theoretical results, which are summarized in next
sections, the ferromagnetic order in Mn-based DMS is mediated by
carriers residing in relatively wide valence bands. In the case of
III-V and II-VI DMS, the holes are coupled to the localised spins
{\em via} a strong, symmetry allowed, antiferromagnetic p-d
interaction. This is in contrast to manganites, where only d
electrons in narrow bands appear to be involved. However, the d
levels of transition metals other than Mn reside in the band gap of
III-V and II-VI compounds. In such a situation the double exchange
may constitute the dominant mechanism of spin-spin interactions.

\section{Magnetic impurities in II-VI and III-V compounds}
\subsection{Energy levels}

We consider tetrahedrally coordinated semiconductors, in which the
magnetic ion occupies the cation sublattice, as found by extended
x-ray absorption fine structure (EXAFS) studies in the case of
Cd$_{1-x}$Mn$_x$Te  \cite{Balz85} and Ga$_{1-x}$Mn$_x$As
\cite{Shio98}. Obviously, magnetic properties of a semiconductor
containing magnetic ion will depend on energetic positions of
states derived from magnetic shells as well as on their
interactions with the host bands. Furthermore, the energy of the
magnetic levels in respect to host bands together with the on-site
correlation energy $U$ determine whether the magnetic ion act as a
dopant and how its charge and spin state depends on the presence of
other impurities.

\begin{figure}
\includegraphics[width=80mm]{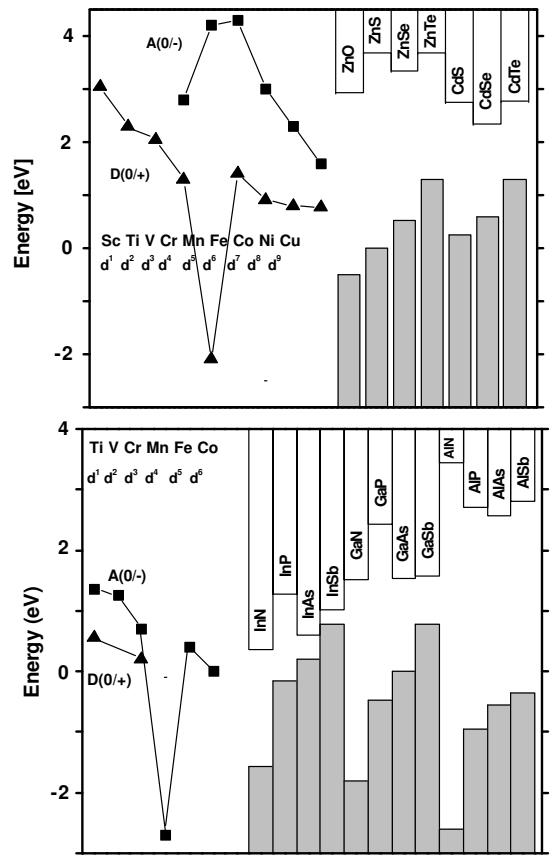}
\caption{Approximate  positions of transition metals levels
relative to the conduction and valence band edges of II-VI (left
panel) and III-V (right panel) compounds. By triangles the
d$^{N}$/d$^{N-1}$ donor and by squares the d$^N$/d$^{N+1}$ acceptor
states are denoted (adapted from Ref.~\cite{Blin02}).}
\label{refrule}
\end{figure}


According to the internal reference rule \cite{Lang88}, the
positions of states derived from magnetic shells do not vary across
the entire family of the II-VI or III-V compounds if the valence
band offsets between different compounds are taken into account. In
Fig.~1, adapted from Ref.~\cite{Blin02}, the data for II-VI and
III-V DMS containing various transition metals are collected. The
symbols D(0/+) and A(0/$-$) denote the donor and acceptor levels
derived from 3d shells of magnetic ions. In the case of II-VI DMS
(left panel), these states correspond to the transformation of the
doubly ionised magnetic ions M$^{2+}$ into M$^{3+}$ and into
M$^{1+}$ ions, in their ground states, respectively, that is to the
lower and upper Hubbard bands. Similarly, in the case of III-V DMS
(right panel) D(0/+) and A(0/-) denote the donor and acceptor
states which, however, in contrast to the situation in II-VI DMS,
correspond to the transformation of the {\em triply} ionised
magnetic ions M$^{3+}$ into M$^{4+}$ and into M$^{2+}$ ions,
respectively. A characteristic evolution of the level energies with
the number of the d electrons is seen in both II-VI and III-V DMS,
the pattern known from atomic spectra but significantly flattened
out in solids by screening and hybridization effects \cite{Zung86}.

It should be noted at this point that the internal reference rule
may serve only for the illustration of chemical trends and not for
extracting the precise values of the ionisation energies. Moreover,
the interaction between the impurity and host states can lead to
the appearance of additional band-gap levels derived from the
semiconductor bands. This may cause some ambiguity concerning the
nature of localised states observed experimentally in these
systems. In particular, a strong p-d hybridization can lead to a
binding of a hole in a Zhang-Rice polaron (charge transfer) state,
which then gives rise to an additional level in the band gap
\cite{Zhan88,Godl92,Beno92,Mizo93,Diet01d}. Furthermore, if the
d$^N$/d$^{N-1}$ donor state resides above the bottom of the
conduction band, the ground state corresponds to a hydrogenic-like
level d$^{N-1}$+e located below the band edge, as observed in
CdSe:Sc \cite{Glod94b}. Similarly, if the acceptor state lies under
the top of the valence band, the ground state corresponds to a
hydrogenic-like acceptor d$^{N+1}$+h, not to the d$^N$ state.
Importantly, band carriers introduce by such magnetic ions can
mediate exchange interactions between the parent spins. Obviously,
energies of hydrogenic-like states follow the band edges, and by no
means are described by the internal reference rule. This appears to
be the situation of the Mn related levels in the gap of III-V
compounds \cite{Diet01d}, the case discussed in detail below.

\subsection{Mn in II-VI compounds}

It is well established that Mn is divalent in II-VI compounds, and
assumes the high spin d$^5$ configuration characterized by $S=5/2$
and $g=2.0$ \cite{Furd88,Koss93,Diet94a}. Indeed, according to
Fig.~1, the Mn ions neither introduce nor bind carriers, but gives
rise to the presence of the localised spin in II-VI DMS. The spin
dependent hybridization between anion p and Mn d states leads to
the superexchange, a short-range antiferromagnetic coupling among
the Mn moments. In order to take the influence of this interaction
into account, it is convenient to parameterize the dependence of
magnetization on the magnetic field in the absence of the carriers,
$M_o(H)$, by the Brillouin function, in which two empirical
parameters, the effective spin concentration
$x_{\mbox{\small{eff}}}N_0 < xN_0$ and temperature
$T_{\mbox{\small{eff}}} > T$, take the presence of the
superexchange interactions into account \cite{Koss93,Diet94a}. The
dependencies $x_{\mbox{\small{eff}}}(x)$ and
$T_{\mbox{\small{AF}}}(x)\equiv T_{\mbox{\small{eff}}}(x) - T$ have
been determined for a number of Mn-based II-VI DMS. Importantly,
the antiferromagnetic superexchange can be overcompensated by
ferromagnetic interactions mediated by band holes \cite{Diet97},
the theoretical prediction confirmed subsequently by the
observation of ferromagnetic ordering in p-type II-VI DMS
\cite{Haur97,Ferr01}.

\subsection{Mn in III-V compounds}

Figure 2, taken from Ref.~\cite{Diet01d}, shows the energetic
position of the Mn impurity level in III-V compounds, as evaluated
by various authors from measurements of optical spectra and
activation energy of conductivity. {\em A priori}, the Mn atom,
when substituting a trivalent metal, may assume either of two
configurations: (i) d$^4$ or (ii) d$^5$ plus a weakly bound hole,
d$^5$+h. Accordingly, the experimentally determined energies
correspond to either d$^4$/d$^5$ or d$^5$+h/d$^5$ levels.

It appears to be a general consensus that the Mn acts as an
effective mass acceptor (d$^5$+h)in the case of antimonides and
arsenides. Such a view is supported by the relatively small Mn
concentrations leading to the insulator-to-metal transition, which
according to the Mott criterion $n^{1/3}a_B = 0.26$, points to a
relatively large extension of the effective Bohr radius $a_B$.
Moreover, the ESR studies of GaAs:Mn reveal, in addition to the
well known spectrum of Mn d$^5$ with the Land\'e factor
$g_{Mn}=2.0$, two additional lines corresponding to $g_1=2.8$ and
$g_2\approx 6$ \cite{Schn87,Mast88,Szcz99b}, which can be described
quantitatively within the $k\cdot p$ scheme for the occupied
acceptor \cite{Schn87,Mast88}. Here, the presence of a strong
antiferromagnetic p-d exchange interaction between the bound hole
and the Mn d-electrons has to be assumed, so that the total
momentum of the complex is $J=1$. In agreement with the model, the
additional ESR lines, in contrast to the $g=2.0$ resonance, are
visible only in a narrow range of the Mn concentration
\cite{Szcz99b}. This range should be greater than the concentration
of compensating donors, and smaller than that at which acceptor
wave functions start to overlap and merge with the valence band.
The antiferromagnetic coupling within the d$^5$+h complex is seen
in a number of experiments, and has been employed to evaluate the
p-d exchange integral $\beta N_0 \approx - 1$~eV \cite{Bhat00} in
GaAs:Mn, the value in agreement with that determined from interband
magnetoabsorption in (Ga,Mn)As \cite{Szcz01}.

\begin{figure}
\includegraphics*[width=90mm]{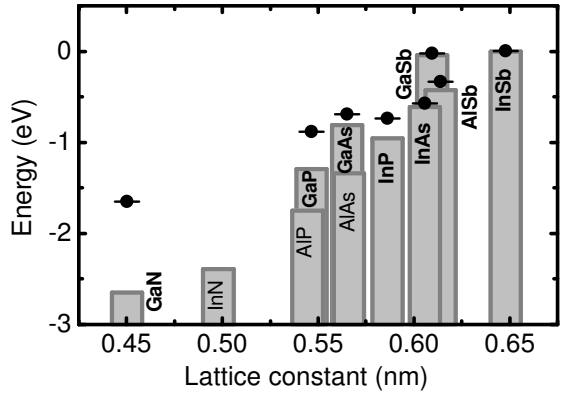}
\caption[]{Experimental energies of Mn levels in the gap of III-V
compounds according to in respect to valence-band edges, whose
relative positions are taken from Ref.~\cite{Vurg01} (after
Ref.~\cite{Diet01d}.}
 \label{fig:III-V}
\end{figure}

Importantly, the above scenario is corroborated by results of
photoemission \cite{Okab98,Okab01} and x-ray magnetic circular
dichroism (XMCD) studies \cite{Ohld00,Ueda01a} in metallic or
nearly metallic (Ga,Mn)As. The latter point to the d$^5$ Mn
configuration. The former are not only consistent with such a
configuration but also lead to the value of $\beta N_0$ similar to
that quoted above, $\beta N_0 \approx - 1.2$~eV \cite{Okab98}.
Furthermore, the photoemission reveals the presence of two features
in the density of states brought about by the Mn constituent: the
original Mn 3d$^5$ states located around 4.5~eV below the Fermi
energy $E_{\mbox{\small{F}}}$, and new states merging with the
valence band in the vicinity of $E_{\mbox{\small{F}}}$
\cite{Okab01}. These new states correspond to acceptors, as
discussed above. They are derived from the valence band by the
Coulomb field as well as by a local Mn potential that leads to a
chemical shift in the standard impurity language, or to a valence
band offset in the alloy nomenclature.

In contrast to antimonides and arsenides, the situation is much
more intricate in the case of phosphides and nitrides. Here, ESR
measurements reveal the presence of a line with $g=2.0$ only
\cite{Mast81,Sun92,Krei96,Zaja01a}, which is thus assigned to d$^5$
centers \cite{Sun92,Krei96,Zaja01a}. Moreover, according to a
detailed study carried out for a compensated n-type GaP:Mn
\cite{Krei96}, the ESR amplitude diminishes under illumination and,
simultaneously, new lines appear, a part of which exhibit
anisotropy consistent with the d$^4$ configuration. This, together
with the apparent lack of evidence for d$^5$+h states, even in
p-type materials, seems to imply that Mn in the ground state
possesses the d$^4$, not d$^5$+h, electron configuration
\cite{Krei96}. This would mean that the Mn energy in Fig.~2 for GaP
\cite{Krei96} and, therefore, for GaN \cite{Wolo01,Blin01} (where
the valence band is lower than in GaP) corresponds to the
d$^4$/d$^5$, not d$^5$+h/d$^5$ level. Such a view appears to be
supported by the {\em ab initio} computation within the local spin
density approximation (LSDA), which points to the presence of the
d-states in the gap of (Ga,Mn)N \cite{Sato01a}. In this situation,
the spin-spin interaction would be driven by a double exchange
mechanism involving hopping of d-electrons \cite{Sato01a,Zene51b},
as in the case of colossal magnetoresistance manganites, not by the
holes in the valence band.

However, the above interpretation has recently been call into
question \cite{Diet01d}. In particular, guided by photoemission
results for II-VI compounds \cite{Mizo93} one expects that the
energy of the d$^4$/d$^5$ level will vary little between arsenides
and nitrides. This implies that this level should reside in the
valence band of GaN despite the 1.8~eV valence band offset between
GaN and GaAs, as shown in Fig.~1. The resulting contradiction with
the LSDA findings can be removed by noting that in the case of
strongly correlated 3d electrons, a semi-empirical LSDA+U approach
is necessary to reconcile the computed and photoemission positions
of states derived from the Mn 3d shell in (Ga,Mn)As
\cite{Okab01,Park00}.

Another important aspect of magnetic acceptors is that the p-d
hybridization, in addition to producing the exchange interaction,
can contribute to the hole binding energy $E_b$. By taking the hole
wave function as a coherent superposition of p-states of anions
adjacent to Mn \cite{Zhan88} and assuming the p-orbitals to be
directed towards the Mn ion, the $T_2$ state has been found to have
30\% lower energy than that corresponding to the mutually parallel
p-orbitals \cite{Diet01d}. This shows that the Kohn-Luttinger
amplitudes away from the $\Gamma$ point of the Brillouin zone are
also involved. In order to evaluated $E_b$, a square-well spherical
potential $V(r) = V_o\Theta(b-r)$ is assumed \cite{Beno92}, whose
depth $V_o$ is determined by the p-d hybridization taking into
account the above mentioned arrangement of the p-orbitals,
\begin{equation}
V_o=\frac{5}{8\pi}\frac{\beta
N_0}{1.04}\left(1-\frac{\Delta_{\mbox{\small{eff}}}}
{U_{\mbox{\small{eff}}}}\right) \left(\frac{a_o}{b}\right)^3.
\end{equation}
Here, the values of $\beta N_0$ are taken from Ref.~\cite{Diet01b};
$\Delta_{\mbox{\small{eff}}}$ is the distance of d$^4$/d$^5$ level
to the top of the valence band, which is evaluated to be 2.7~eV in
(Ga,Mn)As \cite{Okab98}, and is assumed to be reduced in other
compounds by the corresponding valence band offsets, and
$U_{\mbox{\small{eff}}}= 7$~eV is the correlation energy of the 3d
electrons \cite{Okab98,Mizo93}, the energies visible in Figs.~1 and
2. Finally, $b/a_o$ is the well radius in the units of the lattice
constant, and should lie in-between the cation-anion and
cation-cation distance, $\sqrt{3}a_o/4 <b< a_o/\sqrt{2}$. It turns
out that in the case of GaN:Mn the hole is bound by Mn, even in the
absence of the Coulomb potential, $E_b =1.0$~eV for $b = 0.46a_o$.
This demonstrates rather convincingly that a large part of $E_b$
originates indeed from the p-d interaction, indicating that the
Zhang-Rice (ZR) limit \cite{Zhan88,Beno92} is reached in these
systems. One of the important consequences of the the ZR polaron
formation is the shift of the Mott critical concentration towards
rather high values. According to the known relation between $E_b$
and $a_B$ \cite{Berc01a}, the critical hole concentration is
$p_c=4\times 10^{19}$ cm$^{-3}$ in (Ga,Mn)As and at least an order
of magnitude greater in (Ga,Mn)N, if no shallower acceptors are
present.

\section{Theoretical description of Mn-based ferromagnetic
semiconductors}
\subsection{Electronic states and models of carrier-mediated
ferromagnetism}

Despite a considerable effort aiming at elucidating the nature of
ferromagnetism in Mn-based II-VI and III-V DMS, the form of the
relevant minimal Hamiltonian and its universality for all compounds
are still under an active debate. Such a situation reflects the
multifaceted environment, in which the ferromagnetism appears.
Indeed, conceptual and technical difficulties inherent to theory of
strongly correlated and disordered transition-metal compounds are
combined---in ferromagnetic semiconductors---with the intricate
physics of Anderson-Mott localisation and disordered Stoner
magnetism that are specific to heavily doped semiconductors.
Moreover, low-temperature epitaxy, by which III-V materials are
obtained, results in a large concentration of native defects such
as antisites, which act as compensating donors. Another possible
source of local bonds reconstructions is the mechanism of
self-compensation, occurring in heavily doped semiconductors once
the Fermi level reaches the energy triggering defect reactions.
Structural faults may form with neighbor transition metal
impurities defect complexes exhibiting hitherto non-explored
magnetic characteristics. At the same time, strong compensation by
donor-like defects enhances the electrostatic disorder
substantially, leading to deep and long-range potential
fluctuations that result in significant band tailing.

There are two basic approaches to theoretical modeling of the
materials in question: (i) {\it ab initio} or first principles
studies and (ii) theories starting from effective Hamiltonians
containing experimentally determined parameters. Since {\it ab
initio} works are reviewed elsewhere \cite{Kata02}, they will not
be discussed here except for noting that they raise interesting
questions to what extend the local spin density approximation
(LSDA) can handle strong correlation inherent to charge-transfer
and Hubbard-Mott insulators as well as whether the coherent
potential approximation (CPA) can capture the key aspects of the
Anderson-Mott localisation. Our main focus will be on presenting
models employing parametrized Hamiltonians, and on showing their
capabilities and limitations {\it vis \`a vis} experimental
results.

As discussed above, various experiments imply consistently that the
Mn ions are in the high spin 2+ charge state in tetrahedrally
coordinated magnetic semiconductors. Accordingly, the Mn ions are
electrically neutral in II-VI compounds but act as effective mass
acceptors in III-V semiconductors. It is now well established that
in the absence of free carriers the dominant exchange mechanism is
the short-range superexchange in zinc-blende magnetic
semiconductors. This mechanism leads to antiferromagnetic
interactions, except perhaps for some Cr- and V-based compounds,
for which the presence of a ferromagnetic coupling is theoretically
predicted \cite{Blin96a,Blin96b}. Remarkably, owing to the large
exchange energy $|\beta N_0|$ and the high density of states, the
hole-mediated long-range ferromagnetic exchange interaction can
overcome antiferromagnetic superexchange \cite{Diet97}). Indeed, as
already emphasized, the presence of holes is essential for the
existence of the ferromagnetic order in Mn-based semiconductors.
Furthermore, the relevant spin-spin interaction is long range
according to neutron studies \cite{Kepa01}.

It should be recalled at this point that electronic states in doped
semiconductors undergo dramatic changes as a function of the
impurity concentration \cite{Beli94,Edwa95}. Hence, the hole
states, and possibly hole-mediated exchange mechanisms, may {\it a
priori} undergo dramatic changes as function of the Mn content $x$
and the concentrations of acceptors $N_A$ and compensating donors
$N_D$. The evolution of electronic states in doped semiconductors
is governed by the ratio of the average distance between the
carriers $r_c$ to the effective impurity Bohr radius $a_B$,
determined by both Coulomb field and short-range potential of
Eq.~1. In the case of the holes in (Ga,Mn)As, $r_c = (3/4\pi
p)^{1/3}$, $p = xN_0 - N_D$, and $a_B \approx 0.78$ nm
\cite{Berc01a}. A similar value is expected for (Zn,Mn)Te
containing nitrogen acceptors \cite{Ferr01}. However, as already
mentioned, the p-d interaction may significantly contribute to the
impurity binding energy and diminish the effective Bohr radius. In
the range of small impurity concentrations, $r_c \gg a_B$, the
holes are tightly bound to acceptors. Hence, the conductivity
vanishes in the limit of zero temperature. At non-zero
temperatures, the charge transport proceeds either {\em via}
phonon-assisted hopping between occupied and empty acceptors or by
means of thermal activation from the acceptor levels to the valence
band. In a pioneering work Pashitskii and Ryabchenko \cite{Pash79}
evaluated the strength of exchange interactions between localised
spins mediated by band carriers thermally activated from impurity
levels. More recently, Wolff et al.~\cite{Wolf96} considered
carriers localised on impurities and forming bound magnetic
polarons (BMP). It was found that there exists a range of
parameters, in which the coupling between the BMP is ferromagnetic.
This idea was further explored by Bhatt and Wan \cite{Bhat99}, who
examined by Monte Carlo simulations properties of a ferromagnetic
phase transition driven by the interactions between the BMP. In a
more recent work Berciu and Bhatt \cite{Berc01a} discuss, within
the MFA, ferromagnetic interactions mediated by quantum hopping of
holes within the Mn acceptor impurity band in III-V materials. The
compensation is taken into account by assuming that the number of
the holes is smaller than that of the Mn sites, but no effects of
ionised donors on the site energies is taken into account. An
important result is that the disorder in impurity positions tends
to {\em enhance} the magnitude of $T_{\mbox{\small {C}}}$.

Two other groups noted that a long-range exchange interaction
between Mn spins can be mediated by holes undergoing quantum hoping
from the Mn-derived impurity states to the extended valence band
states. Inoue et al.~\cite{Inou00} adopted the Slater-Koster
approach, well known in the physics of resonant states, for the
case of two magnetic impurities. It has been found, by a model
calculation, that the pairs of Mn spins coupled to the valence band
states have a lower energy in the ferromagnetic than in the
antiferromagnetic configuration. Litvinov and Dugayev \cite{Litv01}
suggested than the ferromagnetic spin-spin interaction can
originate from virtual excitations between the acceptor-like
impurity level and the valence band, a variant of the
Bloembergen-Rowland indirect exchange mechanism. They evaluated
Curie temperatures by using a formula, derived originally for
excitations between valence and conduction bands, without proving
its correctness for the case in question.

With the increase of the net acceptor concentration, the impurity
band merges with the valence band. For $r_c \ll a_B$, the holes
reside in the band, and their quasi-free propagation is only
occasionally perturbed by scattering of Mn and other defect
potentials, whose long-range Coulomb part is screened by the
carrier liquid.  Here, the celebrated Ruderman-Kittel-Kasuya-Yosida
(RKKY) mechanism, driven by intraband virtual excitations, is
expected to dominate. In the context of III-V magnetic
semiconductors, this mechanism was discussed by Gummich and da
Cunha Lima \cite{Gumm90} and Matsukura et al.~\cite{Mats98}. At the
same time, the present author and co-workers \cite{Diet97}
demonstrated the equivalence of the RKKY and Zener
\cite{Zene51a,Zene51c} models, at least on the level of the
mean-field and continuous medium approximations. However, with no
doubts, beyond those approximations such equivalence can be
questioned \cite{Seme01}. Within the Zener approach
\cite{Zene51a,Zene51c}, and its nuclear spin variant \cite{Froe40},
the degree of spin ordering, ${\bm M_q}$, at given temperature $T$
is sought by minimizing the total free energy of the spin and
carrier subsystems, $F[{\bm M_q}]$. Here, ${\bm M_q}$ denotes the
Fourier components of localised spin magnetization ${\bm M}({\bm
r})$, so that the minimum of $F[{\bm M_q}]$ for ${\bm M_{q=0}}\ne
0$ implies the ferromagnetic order. Because of its relevance, the
Zener model will be discussed in some details in a subsequent
subsection.

In view of the above discussion a question arises whether the
hole-mediated ferromagnetism appear in the insulator or in the
metallic phase. It is well established that the metal-insulator
transition (MIT) occurs at $r_c \approx 2.4a_B$ in doped
non-magnetic semiconductors \cite{Edwa78}. According to this
criterion one gets the critical hole concentration $p_c = 4\times
10^{19}$ cm$^{-3}$ for $a_B = 0.78$ nm. Experimentally, the MIT
occurs at about 3.5\% of Mn in (Ga,Mn)As, {\it i.e.}, for $N_0x =
7\times 10^{20}$ cm$^{-3}$ \cite{Mats98,Oiwa97,Kats98}. A large
difference between these two values is presumably caused by the
compensation as well as by the enhancement of localisation by the
sp-d exchange scattering \cite{Diet94a}, an effect observed also in
p-(Zn,Mn)Te \cite{Ferr01}. This is documented in both (Ga,Mn)As
\cite{Kats98} and p-(Zn,Mn)Te \cite{Ferr01} by the presence of
negative magnetoresistance and the associated insulator-to-metal
transition driven by the magnetic field \cite{Kats98}. In addition
to the MIT at $x \approx 0.035$ in (Ga,Mn)As, a reentrant insulator
phase is observed for $x > 0.06$ \cite{Mats98}. Presumably, a
self-compensation mechanism is involved, such as the appearance of
interstitial Mn donors, as suggested by first principles studies
\cite{Shir99}. Further efforts are necessary to test this
hypothesis and, more importantly, to push the Mn solubility limit
towards higher $x$ values.

Perhaps, the most intriguing property of the materials in question
is that the ferromagnetism is observed on the both sides of MIT
\cite{Ferr01,Oiwa97,Mats98}. It is, therefore, interesting to
contemplate the nature of electronic states in the vicinity of the
MIT in doped semiconductors. Obviously, the random spatial
distribution of acceptor and donor centers gives rise to strong
spatial fluctuations in the carrier density and states
characteristics. According to the phenomenological two-fluid model
there exist two kinds of relevant states \cite{Paal91}. The first
are strongly localised and thus singly occupied states associated
with the attractive potential of a single majority impurity. The
strongly localised carriers barely contribute to the conduction
process. However, they produce a Curie-like component in the
magnetic susceptibility and give rise to the presence of BMP in
magnetic semiconductors. Obviously, the impurity-like states
dominate deeply in the insulating phase but their presence is
noticeable also in the metallic phase
\cite{Glod94b,Paal91,Glod94a}. The second pool of states determines
the conductivity, so that properties of these states are described
by the scaling theory of MIT. Accordingly, the corresponding
localisation radius $\xi$ is rather controlled by interference of
multi-scattering processes than by the attractive potential of a
single impurity. Thus, $\xi$ of these weakly localised states is
significantly larger than $a_B$, and diverges on approaching the
MIT from the insulator side. It is worth noting that such a
two-fluid model is consistent with a.c. conductivity studies
\cite{Naga01}, which show the coexistence of weakly and strongly
localised states near the MIT in (Ga,Mn)As. Furthermore, the
merging of impurity and band states in this range is substantiated
by angle-resolved photoemission spectra in the same system
\cite{Okab01}.

In order to tell the dominant mechanism accounting for the
existence of long-range spin order in ferromagnetic semiconductors
it is instructive to trace the evolution of their magnetic
properties on crossing the MIT. Remarkably, in contrast to rather
strong changes of resistivity, the evolution of magnetic properties
is gradual. This substantiates the notion that thermodynamic
properties do not exhibit any critical behaviour at MIT as they are
insensitive to large-scale characteristics of the wave functions.
Importantly, the values of Curie temperature are found to grow with
the degree of the material metallicity
\cite{Ferr01,Mats98,Kats01,Pota01}. Moreover, the examination of
magnetization as a function of temperature and magnetic field
indicates that virtually all Mn spins contribute to ferromagnetic
order in the most metallic samples
\cite{Ferr01,Mats98,Oiwa97,Kats01,Pota01}. However, on crossing the
MIT (by lowering $x$), the relative concentration of
ferromagnetically coupled spins decreases substantially. According
to XMCD results \cite{Ohld00}, about 10\% of Mn spins is involved
in ferromagnetism of Ga$_{1-x}$Mn$_x$As with x = 2\%. Also
ferromagnetic resonance studies \cite{Szcz99b} and direct
magnetization measurements demonstrate that only a part of spins
contribute to spontaneous magnetization, while the alignment
process of the remaining moments occurs according to a Brillouin
function for a weakly interacting spin system \cite{Oiwa97}.
Remarkably, the anomalous Hall effect reveals clearly the presence
of the first component but hardly points to the existence of any
loose spins \cite{Ferr01,Mats98}.

The above findings indicate that Mn spins in the regions visited by
itinerant holes are coupled ferromagnetically. These holes set
long-range ferromagnetic correlation between Mn spins, including
those contributing to BMP that are formed around singly occupied
local states. Obviously, the ferromagnetic portion of the material,
and thus the magnitude of spontaneous magnetization, grows with the
dopant concentration, attaining 100\% in the metallic phase. Such a
trend is confirmed by the available data, as discussed above. Thus,
the delocalised or weakly localised holes are responsible for
ferromagnetic correlation in both (Ga,Mn)As and (Zn,Mn)Te:N
\cite{Diet00}. At the same time, mechanisms that involve strongly
localised states, such as excitations from impurity levels or a
direct coupling between BMP, appear to be of lesser importance.

According to the two-fluid scenario referred to above, the BMP are
present on the both sides of the metal-insulator transition (MIT)
\cite{Glod94a}. As already mentioned, the coupling between the BMP
appears to be ferromagnetic \cite{Wolf96}. Since
$T_{\mbox{\small{C}}}$ is proportional to the square of spin vector
length, the weight of the BMP contribution may exceed their
relative concentration.  To gain the Coulomb energy, the BMP are
preferentially formed around close pairs of ionised acceptors. In
the case of III-V materials, one hole localised at two Mn ions
generates, {\em via} Zener's double exchange \cite{Zene51b}, a
strong ferromagnetic coupling that overcompensates the intrinsic
superexchange antiferromagnetic interaction. In contrast, in II-VI
compounds, for which the acceptor cores do not carry any spin and
the degree of compensation by donors is low, BMP are not
preferentially form around the Mn pairs. Accordingly, the pairs of
the close Mn spins remain antiferromagnetically aligned, even in
p-type samples. This diminishes the effective concentration of the
Mn spins contributing to the ferromagnetic order,
$x_{\mbox{\small{eff}}} < x$, and decreases the effective
Curie-Weiss temperature by $T_{\mbox{\small{AF}}}$, both parameters
known from magnetization studies on undoped II-VI DMS. The
weakening of the ferromagnetism by the superexchange in the case of
II-VI compounds, and the virtual absence of the corresponding
effect in III-V materials, where $T_{\mbox{\small{AF}}}\approx 0$
and $x_{\mbox{\small{eff}}} \approx x$, constitutes the important
difference between these two families of magnetic semiconductors.
This fact is taken into account when evaluating the Mn contribution
to the total free energy of particular systems. In worth adding
that recent Monte Carlo simulations \cite{Kene01} carried out
starting from an impurity band model, provides an additional
support for the two fluid scenario.

\subsection{Mean-field Zener model and its limitations}

In terms of the mean-field Zener model of the carrier-mediated
ferromagnetic interactions \cite{Diet97,Zene51c,Lero73,Jung99} the
equilibrium magnetization, and thus the $T_{\mbox{\small{C}}}$ is
determined by the minimum of the Ginzburg-Landau free-energy
functional $F[{\bm M}({\bm r})]$ of the system, where ${\bm M}({\bm
r})$ is the local magnetization of the localised spins. This is a
rather versatile scheme, to which carrier correlation and
confinement \cite{Haur97,Koss00,Diet97,Diet99,Jung99,Lee00},
$k\cdot p$ and spin-orbit couplings
\cite{Ferr01,Diet00,Diet01b,Abol01} as well as weak disorder and
antiferromagnetic interactions \cite{Ferr01,Koss00,Diet97} can be
introduced in a controlled way, and within which the quantitative
comparison of experimental and theoretical results is possible
\cite{Ferr01,Bouk01,Diet01b,Diet01c}. In its general formulation,
the model allows for non-uniform ground states ($M(q>0) \ne 0$),
such spin-density waves or non-collinear (counted) magnetic
structures \cite{Diet99}.

In theory developed by the present authors and co-workers
\cite{Diet97,Diet00,Diet01b}, the hole contribution to $F$ is
computed by diagonalizing the $6\times 6$ Kohn-Luttinger $k\cdot
p$-matrix containing the p-d exchange contribution, and by the
subsequent computation of the partition function $Z$,
$F_{\mbox{\small{c}}} = -k_BT\ln Z$. In the case of a strongly
degenerate Fermi liquid,
|$E_{\mbox{\small{F}}}|/k_{\mbox{\small{B}}}T \gg 1$,
$F_{\mbox{\small{c}}}$ can be replaced by the ground state energy.
However, this approximation would severely overestimate
$T_{\mbox{\small{C}}}$ in materials with low hole densities or with
high $T_{\mbox{\small{C}}}$ values and, therefore, has not been
employed for such situations \cite{Bouk01,Diet00,Diet01b}. The
model is developed for both zinc-blende and wurzite semiconductors,
takes the effects of the spin-orbit interaction into account, and
allows for the presence of both biaxial strain and quantizing
magnetic field. The enhancement of the tendency towards
ferromagnetism by the carrier-carrier exchange interactions is
described by the Fermi liquid parameter $A_{\mbox{\small{F}}}$. The
value $A_{\mbox{\small{F}}} = 1.2$ was evaluated within the LSDA
for the 3D carrier liquid of the relevant density \cite{Jung99}.
Importantly, the formalism has been extended by K\"onig et al.
\cite{Koni00,Koni01a},  who evaluated the dependence of the carrier
free energy on the wave vector $q$. This dependence provides
information on magnetic stiffness, which together with magnetic
anisotropy, determine the dispersion of spin waves \cite{Koni01a}
and the structure of magnetic domains \cite{Diet01c}.

It is, of course, important to comment explicitly main
approximations behind the Zener model. Obviously, this model may
not be applicable to DMS containing magnetic impurities other than
Mn, in which d levels reside in the band gap and correlation energy
$U$ is relatively small. In the Mn-based DMS, however, the magnetic
constituent appears to be well described in terms of the the
Anderson impurity model and, thus, such DMS can be regarded as
charge transfer insulators. This means that the spins are
localised, so that the d electrons do not participate in charge
transport. There exist, however, quantum fluctuations
(hybridization) between the p and d orbitals, which result in the
Kondo-like interaction that couples the spin and the carrier parts
of the free energy. It should be emphasized that results of the
LSDA computations \cite{Kata02} point to a rather large weight of
the d electrons at the Fermi level and, hence, imply the
ferromagnetism to be driven by the mechanism of double exchange.
This controversy is perhaps the most intriguing open issue in
theory of magnetic semiconductors.

The use of the spin, not carrier magnetization, as the order
parameter in the Zener model is a consequence of the adiabatic
approximation: spin dynamics is assumed to be much slower than that
of the carriers. This approximation may break down if, for
instance, a characteristic energy of spin-spin antiferromagnetic
interactions would become greater than $T_{\mbox{\small{C}}}$.
Furthermore, the order parameter can be regarded as continuous as
long as the concentration of the spins is much larger than that of
the holes, $x_{\mbox{\small{eff}}}N_0 \gg p$. Importantly, in this
range, the mean-field value of the ordering temperature
$T_{\mbox{\small{C}}}({\bm q})$ deduced from the Zener and the RKKY
model are identical, independently of microscopic spin distribution
\cite{Diet97}.

However, in the opposite limit, $x_{\mbox{\small{eff}}}N_0 < p$,
important changes in the hole response function occur at the length
scale of a mean distance between the localised spins. Accordingly,
the description of spin magnetization by the continuous-medium
approximation, which constitutes the basis of the Zener model,
ceases to be valid. In contrast, the RKKY model is a good starting
point in this regime \cite{Ferr01}, as it provides the dependence
of the interaction energy of particular spin pairs as a function of
their distance. This makes it possible to evaluate the system
energy for a given distribution of the localised spins. The
resulting competition between the ferromagnetic and
antiferromagnetic interactions is expected to grow with
$p/x_{\mbox{\small{eff}}}N_0$, and may lead, through non-collinear
(counted) spin arrangement \cite{Koni01b,Zara01}, to the spin-glass
phase \cite{Egge95}. Alternatively, with decreasing $x$ at given
$p$, the Kondo effect may show up \cite{Diet97}.

It is interesting to note that the role of thermodynamic
fluctuations of magnetization, that is the inaccuracy of the
mean-field approximation (MFA), grows also with
$p/x_{\mbox{\small{eff}}}N_0$. It is well known that the MFA
results are exact, also in low-dimensional systems, if the range of
ferromagnetic interactions is infinite \cite{Fish72}. The decay of
the strength of the carrier-mediated exchange interaction with the
distance $r$ between two Mn spins is described by the RKKY
oscillatory function. At small $r$, the interaction is
ferromagnetic, and then changes sign at $r = 1.2r_c$, where $r_c$
is an average distance between the carriers that mediate the
spin-spin coupling. This implies that the MFA is valid
quantitatively at $p << x_{\mbox{\small{eff}}}N_0$, a conclusion
consistent with the estimate of $T_{\mbox{\small{C}}}$ taking the
spin wave excitations into account \cite{Koni00}. Actually,
however, the range of validity of the MFA is significantly larger
\cite{Koni01b} than that initially proposed \cite{Koni00,Schl01a},
as the magnitudes of spin stiffness evaluated within the 6x6
Luttinger model is much greater \cite{Koni01a} than those obtained
for a simple parabolic band \cite{Koni00}. The dynamic mean-field
approach \cite{Chat01} and, especially, hybrid Monte-Carlo
algorithms \cite{Schl01b} have potential to sheet some light on
ferromagnetism in the regime beyond the validity of the MFA.

Finally, we address the question about the role played by disorder.
Since thermodynamic properties of the carrier liquid are relatively
weakly perturbed by scattering, its effect on the hole-mediated
ferromagnetism can be neglected to a first approximation. When
disorder grows and the MIT is approached, the mean free path
becomes comparable to the inverse Fermi wave vector.  Within the
Zener model, the effect of the finite mean free path $l_{e}$ can be
described by scattering broadening of the density of states
\cite{Koss00,Diet97}, which reduces $T_{\mbox{\small{C}}}$.
Technically, this follows from the averaging of the free energy
over possible impurity distributions. Equivalently, diffusive
character of carrier transport leads to exponential dumping of the
RKKY interactions at distances longer than $l_{e}$. However, large
fluctuations in the carrier distribution have to be taken into
account at criticality and on the insulator side of the MIT. As
already argued, the magnetization is small in areas which are not
visited by the carriers, whereas its magnitude is large there,
where delocalised or weakly localised carriers reside. Remarkably,
this enhances $T_{\mbox{\small{C}}}$ over the value expected for
the average carrier density \cite{Berc01a} but reduces the
magnitude of sample-average spontaneous magnetization
\cite{Ferr01,Diet00,Diet01b}. The interplay between Anderson-Mott
localisation, Stoner magnetism, and carrier-mediated spin-spin
interaction is certainly an appealing area for future research.
Recent theoretical works in this direction are indeed encouraging
\cite{Chud01}.

\section{Comparison of the Zener model to selected experimental results}
\subsection{Magnetic circular dichroism in (Ga,Mn)As}

Within the Zener model, the strength of the ferromagnetic spin-spin
interaction is controlled by the $k\cdot p$ parameters of the host
semiconductor and by the magnitude of the spin-dependent coupling
between the effective mass carriers and localised spins. In the
case of II-VI DMS, detailed information on the exchange-induced
spin-splitting of the bands, and thus on the coupling between the
effective mass electrons and the localised spins has been obtained
from magnetooptical studies \cite{Koss93,Diet94a}. A similar work
on (Ga,Mn)As \cite{Ando98,Szcz99a,Besc99} led to a number of
surprises. The most striking was the opposite order of the
absorption edges corresponding to the two circular photon
polarizations in (Ga,Mn)As comparing to II-VI materials. This
behaviour of circular magnetic dichroism (MCD) suggested the
opposite order of the exchange-split spin subbands, and thus a
different origin of the sp-d interaction in these two families of
DMS. A new light on the issue was shed by studies of
photoluminescence (PL) and its excitation spectra (PLE) in p-type
(Cd,Mn)Te quantum wells \cite{Haur97}.  As shown schematically in
Fig.~3, the reversal of the order of PLE edges corresponding to the
two circular polarizations results from the Moss-Burstein effect,
that is from the shifts of the absorption edges associated with the
empty portion of the valence subbands in the p-type material. This
model was subsequently applied to interpret qualitatively the
magnetoabsorption data for metallic (Ga,Mn)As \cite{Szcz99a}. More
recently, the theory was extended by taking into account the effect
of scattering-induced mixing of $k$ states \cite{Szcz01}. As shown
in Fig.~4, this approach explains the slop of the absorption edge
as well as its field-induced splitting assuming the value of the
p-d exchange energy $\beta N_0 = -1$~eV.

\begin{figure}
\includegraphics*[width=90mm]{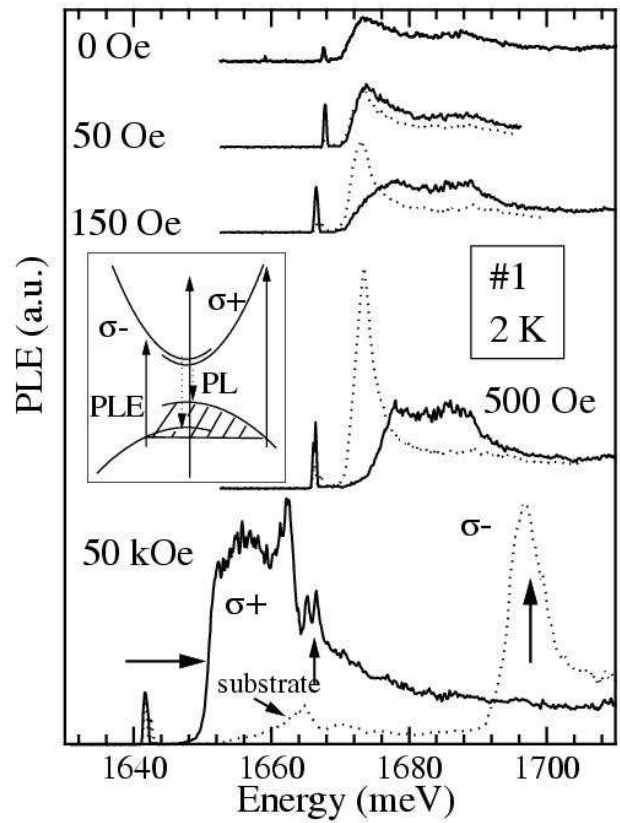}
 \caption{Photoluminescence excitation spectra (PLE), that is the
photoluminescence (PL) intensity as a function of the excitation
photon energy intensity, for $\sigma^+$ (solid lines) and
$\sigma^-$ (dotted lines) circular polarizations at selected values
of the magnetic field in a modulation-doped p-type quantum well of
Cd$_{0.976}$Mn$_{0.024}$Te at 2 K. The photoluminescence was
collected in $\sigma^+$ polarization at energies marked by the
narrowest features. The sharp maximum (vertical arrow) and
step-like form (horizontal arrow) correspond to quasi-free exciton
and transitions starting at the Fermi level, respectively. Note
reverse ordering of transition energies at $\sigma^+$ and
$\sigma^-$  for PL and PLE (the latter is equivalent to optical
absorption). The band arrangement at 150 Oe is sketched in the
inset (after Ref.~\cite{Haur97}).}
\end{figure}

\begin{figure}
\includegraphics*[width=80mm]{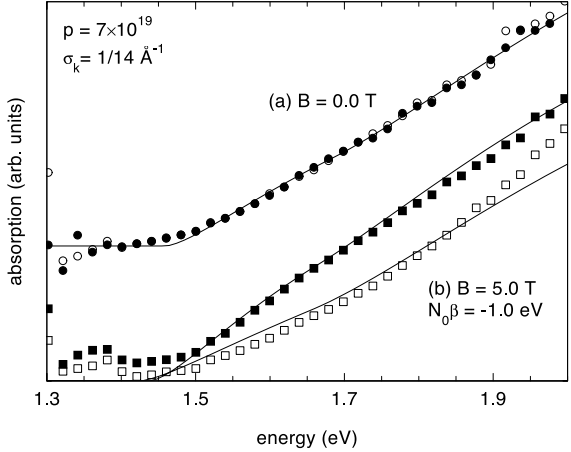}
\caption{Transmission of Ga$_{0.968}$Mn$_{0.032}$As film for two
circular light polarizations in the Faraday configuration in the
absence of the magnetic field (data shifted up for clarity) and in
5~T at 2~K (points) \cite{Szcz99a}. Solid lines are calculated for
the hole concentration $p=7\times 10^{19}$ cm$^{-3}$, exchange
energy $N_0\beta=-1$~eV, and allowing for scattering-induced
breaking of the $k$ selection rules \cite{Szcz01}.}
\end{figure}

Surprisingly, however, the anomalous sign of the MCD was present
also in non-metallic (Ga,Mn)As, in which EPR signal from occupied
Mn acceptors was seen \cite{Szcz99b}. It has, therefore, been
suggested that the exchange interaction between photo- and
bound-holes is responsible for the anomalous sign of the MCD in
those cases \cite{Szcz99a}. The presence of such a strong exchange
mechanism is rather puzzling, and it should be seen in non-magnetic
p-type semiconductors. At the same time, according to our two-fluid
model, the co-existence of strongly and weakly localised holes is
actually expected on the both sides of the MIT. Since the
Moss-Burstein effect operates for interband optical transitions
involving weakly localised states, it leads to the sign reversal of
the MCD, also on the insulating side of the MIT.

Another striking property of the MCD is a different temperature
dependence of the normalized MCD at low and high photon energies in
ferromagnetic (Ga,Mn)As \cite{Besc99}. This observation was taken
as an evidence for the presence of two spectrally distinct
contributions to optical absorption \cite{Besc99}. A quantitative
computation of MCD spectra has recently been undertaken
\cite{Diet01b}. The theoretical results demonstrate that because of
the Moss-Burstein effect, the magnetization-induced splitting of
the bands leads to a large energy difference between the positions
of the absorption edges corresponding to the two opposite circular
polarizations. This causes an unusual dependence of the low-energy
onset of MCD on magnetization, and thus on temperature. These
considerations lead to a quantitative agreement with the
experimental findings, provided that the actual hole dispersion and
wave functions are taken for the computation of MCD.

In conclusion, a.c. conductivity  in the far infrared \cite{Naga01}
as well as to photoemission \cite{Okab98,Okab01} and XMCD
\cite{Ohld00,Ueda01a} in the range of high photon energies,
together with magnetooptical characteristics discussed here, are
consistent with the picture of electronic states advocated for
(Ga,Mn)As in the previous section. In particular, (Ga,Mn)As
exhibits properties generic to doped semiconductors in the vicinity
of the metal-insulator trsansition. Furthermore, no experimental
results have so far been collected in favor of the valence band
splitting corresponding to the p-d exchange integral as high as
$\beta N_0 \approx -4$~eV, as suggested by {\em ab initio}
computations within the LSDA \cite{Kata02}.

\subsection{Curie temperature and spontaneous magnetization in
p-type and n-type Mn-based DMS}

In order to compare theoretical expectations concerning
$T_{\mbox{\small{C}}}$ to experimental results, it is convenient to
introduce the normalized ferromagnetic temperature
$T_{\mbox{\small{F}}}/x_{\mbox{\small{eff}}}= (T_{\mbox{\small{C}}}
+ T_{\mbox{\small{AF}}})/x_{\mbox{\small{eff}}}$ which, within the
mean-field Zener model, should not depend on the Mn concentration
$x$.  We recall that $x_{\mbox{\small{eff}}}<x$ and
$T_{\mbox{\small{AF}}}>0$ take into account the presence of
antiferromagnetic exchange interaction in II-VI DMS. Figure 5
presents $T_{\mbox{\small{F}}}/x_{\mbox{\small{eff}}}$ for
Ga$_{1-x}$Mn$_x$As \cite{Ohno96a,Mats98,Omiy00},
p-Zn$_{1-x}$Mn$_{x}$Te \cite{Ferr01,Andr01}, and quantum wells of
p-Cd$_{1-x}$Mn$_x$Te \cite{Koss00,Cibe02} as a function the Fermi
wavevector determined from the value of the hole concentration $p$
assuming the Fermi sphere to be isotropic. The hole concentration
was deduced either from the Hall resistance \cite{Ferr01,Omiy00} or
form the Moss-Burstein shift \cite{Koss00,Bouk01}. We have to
emphasize that because of a contribution from the anomalous (spin)
Hall effect and a non-uniform hole distribution in thin layers, the
evaluation of the hole concentration is by no means straightforward
in magnetic semiconductors.

\begin{figure}
\includegraphics*[width=90mm]{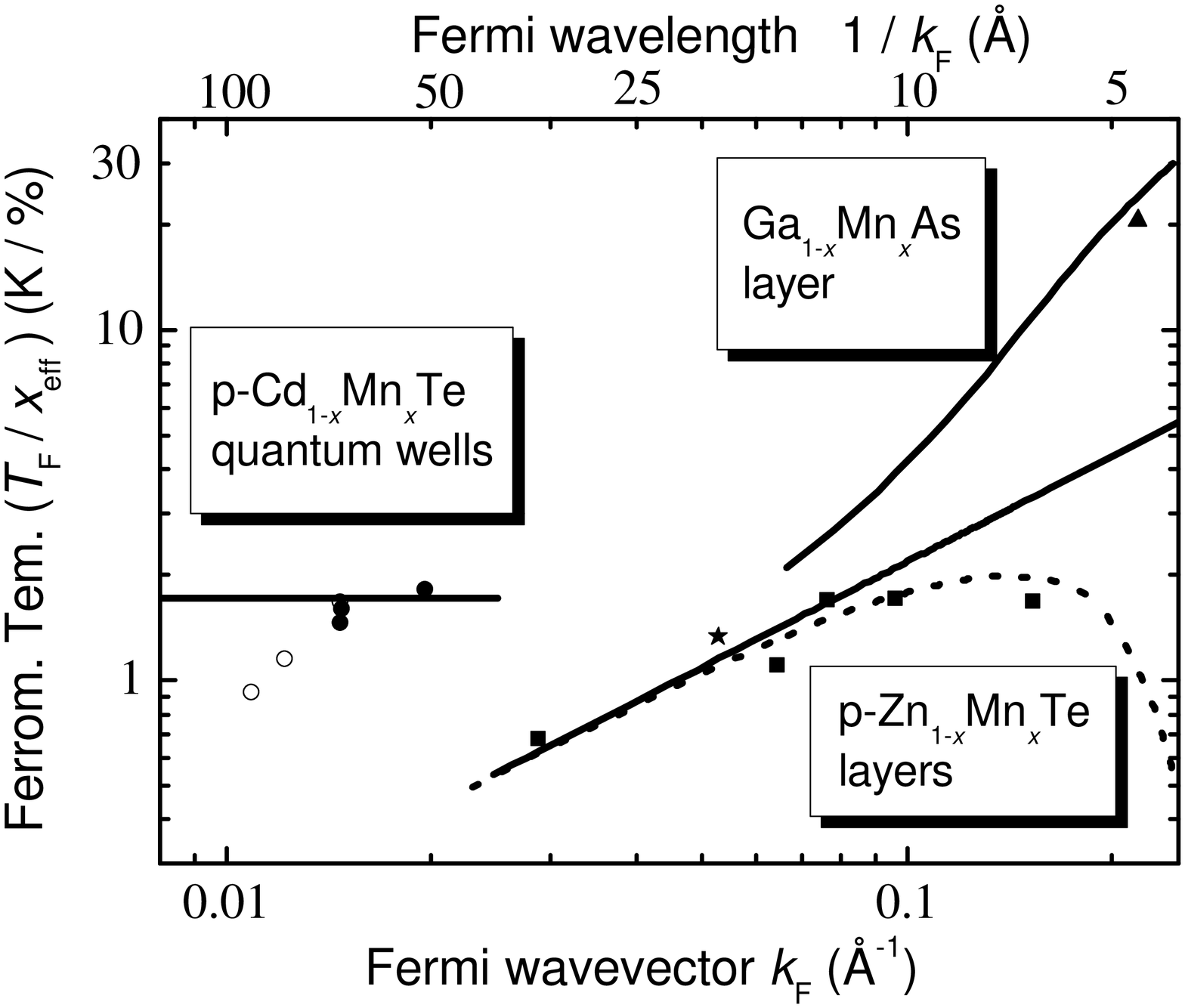}
\caption[]{Experimental (symbols) and calculated (lines) normalized
ferromagnetic temperature, $T_{\mbox{\small F}}/10^2x_{\mbox{\small
eff}}$, versus the wave vector at the Fermi level for
Ga$_{1-x}$Mn$_x$As (triangle) \cite{Mats98,Omiy00},
Zn$_{1-x}$Mn$_x$Te:N (squares) \cite{Ferr01,Andr01},
Zn$_{1-x}$Mn$_x$Te:P (star) \cite{Andr01}, and quantum well of
p-Cd$_{1-x}$Mn$_x$Te (circles) \cite{Koss00,Cibe02}.  Solid lines:
Zener and $6\times6$ Luttinger model for the 3D  \cite{Diet01b})
and 2D case \cite{Diet97}; dotted line: the RKKY and $6\times6$
Luttinger model for $x_{\mbox{\small eff}}=0.015$, taking into
account the effect of the antiferromagnetic interactions on
statistical distribution of unpaired Mn spins \cite{Ferr01}.}
\label{fig:znmnte}
\end{figure}

A number of important conclusions emerges from the comparison of
experimental and theoretical results. First, the theory
\cite{Diet00,Diet01b} with $\beta N_0 = -1.2$~eV \cite{Okab98}
explains the large magnitude of $T_{\mbox{\small{C}}}$ = 110~K
\cite{Ohno96a,Mats98} for Ga$_{0.947}$Mn$_{0.053}$As containing
$3.5\times 10^{20}$ holes per cm$^3$ \cite{Omiy00}. Second, in
contrast to Mn-based III-V DMS, it is essential---in the case of
II-VI materials---to take into account the presence of magnetically
inert nearest-neighbor Mn pairs.  Such pairs not only lowers the
effective Mn concentration but also make the antiferromagnetic
portion of the RKKY interaction to become more significant, which
lowers $T_{\mbox{\small{C}}}$ at large $p$ (dotted line)
\cite{Ferr01}. Third, scaling theory of electronic states near the
MIT, discussed in the previous section, makes it possible to
explain the presence of ferromagnetism on the both sides of the
MIT, and a non-critical evolution of $T_{\mbox{\small{C}}}$ across
the critical point, a behaviour observed in both (Ga,Mn)As
\cite{Mats98} and p-(Zn,Mn)Te \cite{Ferr01,Andr01,Sawi02}.
Importantly, in agreement with this scenario, a ferromagnetic
component of the material increases with the hole concentration
\cite{Ferr01}. However, as already explained, because of RKKY
oscillations, the present theory (solid lines) overestimates the
magnitude of $T_{\mbox{\small{C}}}$ if the hole and Mn
concentrations becomes comparable, $p \approx
x_{\mbox{\small{eff}}}N_0$. In contrast, the actual value of
$T_{\mbox{\small{C}}}$ can be larger than that of Fig.~5 on the
insulator side of the MIT because of the fluctuations in the
carrier density and a non-linear contribution of spin clusters
within the BMP. This might be the reason for a relatively high
values of $T_{\mbox{\small{C}}}$ detected recently in (In,Mn)As
\cite{Oiwa01}. Last but not least, according to Fig.~5, in
modulation doped (Cd,Mn)Te/(Cd,Mg,Zn)Te:N heterostructures, due to
the enhanced density-of-states (DOS) at low energies in 2D systems
and reduced localisation by the distant nitrogen acceptors, the
ferromagnetic transition occurs even for the densities of the hole
liquid as low as $10^{17}$ cm$^{-3}$ \cite{Haur97,Koss00}. The
present theory describes correctly the dependence of
$T_{\mbox{\small{C}}}$ on $x$ and $p$, provided that the exchange
interaction between the holes and disorder broadening of DOS are
taken into account \cite{Koss00,Bouk01}.

Two effects appear to account for the greater
$T_{\mbox{\small{C}}}$ values in p-(Ga,Mn)As than in p-(Zn,Mn)Te at
given $p$ and $x$. First is the smaller magnitude of the spin-orbit
splitting between the $\Gamma_8$ and $\Gamma_7$ bands in arsenides,
$\Delta_o = 0.34$~eV, in comparison to that of tellurides,
$\Delta_o = 0.91$~eV. Once the Fermi energy $E_{\mbox{\small{F}}}$
approaches the $\Gamma_7$ band, the density-of-states effective
mass increases, and the reduction of the carrier spin
susceptibility by the spin-orbit interaction is diminished. The
computed value of $T_{\mbox{\small{C}}}$ for $p = 3\times
10^{20}$~cm$^{-3}$ is greater by a factor of four in (Ga,Mn)As than
that evaluated in the limit $\Delta_o >> E_{\mbox{\small{F}}}$. The
other difference between the two materials is the destructive
effect of antiferromagnetic interactions, which operate in II-VI
compounds but are of minor importance in III-V materials, as argued
in the previous section.

Since in semiconductors the magnitude of exchange splitting of
bands is comparable to the Fermi energy, the growth of spontaneous
magnetization $M_s$ on lowering temperature deviates from the
Brillouin-type behaviour even in the mean-filed approximation
\cite{Haur97,Ferr01,Diet97,Diet01b,Koss00}. Because of extremely
low hole densities, the effect is particularly well visible in the
case of modulation-doped (Cd,Mn)Te/(Cd,Mg,Zn)Te:N heterostructures
\cite{Haur97,Koss00}. Furthermore, the contribution from carrier
magnetic moments and disorder-induced fluctuations in the carrier
distribution constitute other sources of corrections to the
Brillouin-type dependence. However, with the increase of the hole
concentration, $M_s(T)$ evaluated within the mean-field Zener model
tends toward the Brillouin function for material parameters of
(Ga,Mn)As \cite{Diet01b}, an expectation corroborated by the
experimental results \cite{Mats98,Pota01}.

Finally, we turn to materials, in which the Fermi level resides in
the s-type conduction band.  Because of small s-d exchange energy
and low density of states, no ferromagnetism is expected above 1~K
in such a case \cite{Diet97}. Experimental results for n-(In,Mn)As
\cite{Moln91}, (Ga,Mn)As:Sn \cite{Sato01b}, and (Zn,Mn)O:Al
\cite{Andr01} confirm this expectation. With this in mind, the
presence of indications of ferromagnetism in n-type (Ga,Mn)N
\cite{Over01a} is challenging.

\subsection{Effects of strain}

Both hydrostatic and axial strain affect the valence band, and thus
alter the magnitude of the density of states and
$T_{\mbox{\small{C}}}$. Quantitatively, however, the effect is
evaluated to be small \cite{Diet01b}. There exists another
mechanism by which strain may affect $T_{\mbox{\small{C}}}$. It is
presently well known that the upper limit of the achievable carrier
concentration is controlled by pinning of the Fermi level by
impurity or defect states in virtually all compound semiconductors.
Since the energies of such states in respect to bands vary strongly
with the bond length, the hole concentration and thus
$T_{\mbox{\small{C}}}$ will depend on strain.

Apart from $T_{\mbox{\small{C}}}$ and spontaneous magnetization
$M_s$, it is interesting to consider means making it possible to
tailor magnetic anisotropy, and thus the direction of the
spontaneous magnetization, the coercive force, the switching field,
the domain structure. Already early studies of the ferromagnetic
phase in In$_{1-x}$Mn$_x$As \cite{Mune93} and Ga$_{1-x}$Mn$_x$As
\cite{Ohno96b,Shen97} demonstrated the existence of sizable
magnetic anisotropy. Magnetic anisotropy is usually associated with
the interaction between spin and orbital degrees of freedom of the
d-electrons. According to the model advocated here, these electrons
are in the d$^5$ configuration. For such a case the orbital
momentum $L = 0$, so that effects stemming from the spin-orbit
coupling are expected to be rather weak. It has, however, been
noted that the interaction between the localised spins is mediated
by the holes that have a non-zero orbital momentum \cite{Diet00}.
An important aspect of the Zener model is that it does take into
account the anisotropy of the carrier-mediated exchange interaction
associated with the spin-orbit coupling in the host material
\cite{Diet00,Diet01b,Abol01}, an effect difficult to include within
the standard approach to the RKKY interaction \cite{Zara01}.

A detail numerical analysis of anisotropy energies has been carried
out for a number of experimentally important cases
\cite{Diet00,Diet01b,Abol01}. In particular, the cubic anisotropy
as well as uniaxial anisotropy under biaxial strain have been
studied as a function of the hole concentration $p$. The
computation indicates that for the parameters of Ga$_{1-x}$Mn$_x$As
films grown along the [001] direction, the spontaneous
magnetization ${\bm M}$ lies in the (001) plane, and the easy axis
is directed along [100] or along [110] (or equivalent) crystal axis
depending on the degree of the occupation the hole subbands as well
as on their mixing by the p-d and $k\cdot p$ interactions. As a
result, the easy axis fluctuates between [100] and [110] as a
function of $p$, the preferred direction for typical hole
concentrations being [110].  The magnitude of the external magnetic
field $H_{cu}$ that aligns ${\bf M}$ along the hard direction in
the (001) plane is evaluated to be up to 0.2 T \cite{Diet01b}.
Since, however, the orientation of the easy axis changes rapidly
with $p$ and $M$, disorder--which leads to broadening of hole
subbands--will presumably diminish the actual magnitude of magnetic
anisotropy. The field $\mu_oH_{cu}$ determines also the magnitude
of the switching field, which could be observed in microstructures
containing only a single domain. In macroscopic films, however,
smaller values of the coercive field $\mu_oH_c$ are expected as
actually observed: typically $\mu_oH_c = 4$~mT for the magnetic
field along the easy axis in the (001) plane in Ga$_{1-x}$Mn$_x$As
\cite{Shen97}.

It can be expected that strain engineering can efficiently control
magnetic properties resulting from the hole-mediated exchange.
Indeed, sizable lattice-mismatch driven by biaxial strain is known
to exist in semiconductor layers. In some cases, particularly if
epitaxy occurs at appropriately low temperatures, such strain can
persist even beyond the critical thickness due to relatively high
barriers for the formation of misfit dislocations. It has been
found that the biaxial strain leads to uniaxial anisotropy, whose
magnitude can be much greater than that resulting from either cubic
anisotropy or stray fields. As shown in Fig.~6, for the
experimentally relevant values of $p$ and $M$, the easy axis is
predicted to be oriented along [001] direction for the tensile
strain, whereas it should reside in the (001) plane for the case of
unstrained or compressively strained films
\cite{Diet00,Diet01b,Abol01}. This is corroborated by the
experimental study \cite{Ohno96b,Shen97}, in which either (In,Ga)As
or GaAs substrate was employed to impose tensile or compressive
strain in Ga$_{1-x}$Mn$_x$As, respectively. In particular, for the
Ga$_{0.965}$Mn$_{0.035}$As film on GaAs, for which $\epsilon_{xx} =
-0.24$\%, the anisotropy field $\mu_oH_{un} = 0.4 \pm 0.1$~T is
observed \cite{Ohno96b,Shen97}, in quantitative agreement with the
theoretical results of Fig.~6. This field is about two orders of
magnitude greater than that evaluated from the extrapolation of ESR
data on single-ion anisotropy at low $x$ \cite{Fedo01}, a result
confirming the dominant contribution of the holes to the magnitude
of $H_{un}$. Though no theoretical computations have been performed
for In$_{1-x}$Mn$_x$As, the qualitatively similar effect of biaxial
strain is expected, in agreement with the early experimental
results \cite{Mune93}.

\begin{figure}
\includegraphics*[width=90mm]{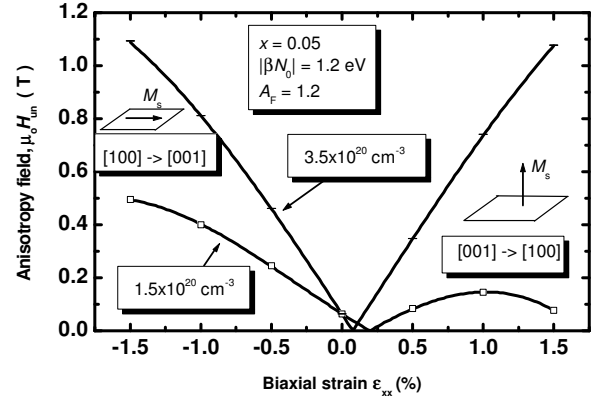}
\caption[]{Computed minimum magnetic field $H_{un}$ necessary to
align the saturation value of magnetization $M_s$ along the hard
axis as a function of biaxial strain component $\epsilon_{xx}$ for
two values of the hole concentrations in Ga$_{0.95}$Mn$_{0.05}$As
\cite{Diet01b}. The symbol [100] $\rightarrow$ [001] means that the
easy axis is along [100], so that $H_{un}$ is applied along [001].}
\label{fig:bunvse_11}
\end{figure}

It worth noting that similarly to strain, also confinement of the
holes affects the magnetic anisotropy –-- in accord with the
theoretical model, the easy axis is oriented along the growth
direction in the ferromagnetic p-(Cd,Mn)Te quantum wells
\cite{Haur97,Koss00}.

\subsection{Domain structure}

Recently, the structure of magnetic domains in Ga$_{1-x}$Mn$_x$As
under tensile strain has been determined by micro-Hall probe
imaging \cite{Shon00}. The regions with magnetization oriented
along the [001] and [00 ] easy axis form alternating stripes
extending in the [110] direction. This indicates, for either Bloch
or N\'eel domain walls, that the in-plane easy axis is rather along
[110] than along [100] directions, a conclusion consistent with the
theoretical expectation for in-plane (cubic) magnetic anisotropy
presented above. As shown in Fig.~7, the experimentally determined
stripe width is $W = 1.5$~$\mu$m at 5~K for 0.2~$\mu$ film of
Ga$_{0.957}$Mn$_{0.043}$As on Ga$_{0.84}$In$_{0.16}$As, for which
tensile strain of $\epsilon_{xx} = 0.9$\% is expected.

\begin{figure}
\includegraphics*[width=90mm]{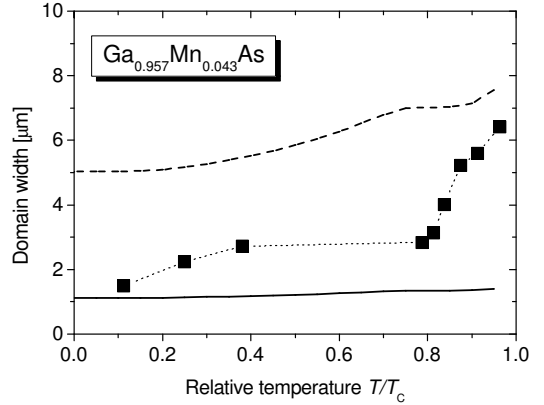}
\caption[]{Temperature dependence of the width of domain stripes as
measured for the Ga0.957Mn0.043As film with the easy axis along the
growth direction (full squares) \cite{Shon00}. Computed domain
width is shown by the solid line. The dashed line is computed
assuming that the parameter $\lambda_c$ (the ratio of the domain
wall and stray field energies) is by a factor of 1.8 greater
\cite{Diet01c}.}
\end{figure}

According to micromagnetic theory \cite{Hube98}, $W$ is determined
by the dimensionless parameter $\lambda_c$, which is given by the
ratio of the domain wall and stray field energies. The former is
proportional to $M^2$, whereas the latter scales with the product
of anisotropy energy $K_u$ \cite{Diet01b,Abol01} and magnetic
stiffness $A$ \cite{Koni01a}. Figure 7 presents the calculated
values of $W(T)$ \cite{Diet01c} in comparison to the experimental
data for (Ga,Mn)As \cite{Shon00}. The material parameters collected
in Ref.~\cite{Diet01b}, and employed to generate theoretical
results of Figs.~5 and 6, have been adopted. Furthermore, in order
to establish the sensitivity of the theoretical results to the
parameter values, the results calculated for a value of $\lambda_c$
1.8 times larger are included as well. The computed value for low
temperatures, $W = 1.1$~$\mu$m, compares favorably with the
experimental finding, $W = 1.5$~$\mu$m. However, the model predicts
much weaker temperature dependence of the domain width $W$ than
observed experimentally, which are linked \cite{Diet01c} to
critical fluctuations, disregarded in the mean-field approach.

\section{Towards high-temperature ferromagnetic semiconductors}
\subsection{Mn-based tetrahedrally  coordinated DMS}

In view of the general agreement between experiment and theory for
$T_{\mbox {\small C}}$ and the magnetic anisotropy, it is tempting
to extend the model for material systems that might be suitable for
fabrication of functional ferromagnetic semiconductors. For
instance, the model suggests immediately that $T_{\mbox {\small
C}}$ values above 300~K could be achieved in
Ga$_{0.9}$Mn$_{0.1}$As, if such a large value of $x$ would be
accompanied by a corresponding increase in the hole concentration.
Figure 8 presents the values of $T_{\mbox {\small C}}$ computed for
various tetrahedrally coordinated semiconductors containing 5\% of
Mn per cation and $3.5\times 10^{20}$ holes per cm$^3$
\cite{Diet00}. In addition to adopting the tabulated values of the
band structure parameters, the same value of $\beta =
\beta$[Ga$_{1-x}$Mn$_x$As] for all group IV and III-V compounds was
assumed, which results in an increase of $|\beta N_0| \sim
a_o^{-3}$, where $a_o$ is the lattice constant, a trend known to be
obeyed within the II-VI family of magnetic semiconductors. For the
employed parameters, the magnitude of $T_{\mbox {\small C}}$ for
the cubic GaN is by 6\% greater than that computed for the wurzite
structure.

\begin{figure}
\includegraphics*[width=90mm]{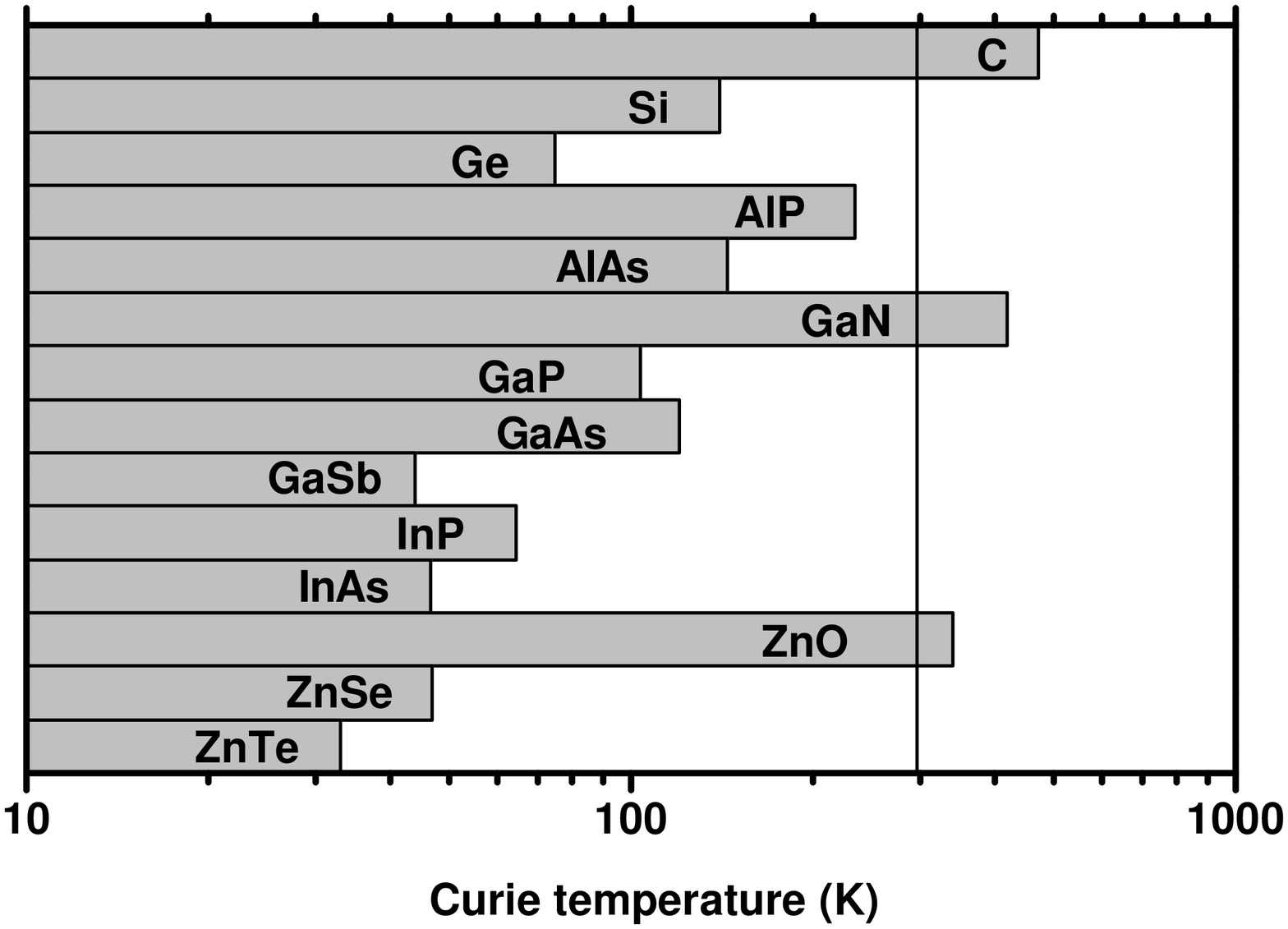}
\caption[]{Computed values of the Curie temperature
$T_{\mbox{\small C}}$ for various p-type semiconductors containing
5\% of Mn per cation (2.5\% per atom) and $3.5\times 10^{20}$ holes
per cm$^{3}$ (after Refs.~\cite{Diet00,Diet01b}).}
\label{fig:trends}
\end{figure}

The data (Fig.~8) demonstrate that there is much room for a further
increase of $T_{\mbox {\small C}}$ in p-type magnetic
semiconductors. In particular, a general tendency for greater
$T_{\mbox {\small C}}$ values in the case of lighter elements stems
from the corresponding increase in p-d hybridization and the
reduction of the spin-orbit coupling. It can be expected that this
tendency is not altered by the uncertainties in the values of the
relevant parameters. Indeed, the results of Fig.~8 have triggered a
considerable fabrication effort, which is bringing a number of
striking developments. In particular Ge$_{1-x}$Mn$_x$ was found to
be p-type and to exhibit $T_{\mbox {\small C}}$ in the excess of
100~K \cite{Park01}. An even higher magnitude of $T_{\mbox {\small
C}}$, in excess of 200~K, was found in (Ga,Mn)P:C \cite{Over01b}, a
value consistent with the prediction of Fig.~8
\cite{Diet00,Diet01b} for the employed Mn content $x=9.4$\%
\cite{Over01b}. Moreover, there are indications of ferromagnetism
in (Ga,Mn)N \cite{Kuwa01,Theo01,Sono01,Reed01}, in some cases with
$T_{\mbox{\small {C}}}$ near or above 300~K
\cite{Theo01,Sono01,Reed01}. While further experimental work is
needed to elucidate a possible role of precipitates of various
Mn-Ga and Mn-N compounds, the highest value suggested up to now,
$T_{\mbox{\small {C}}} = 940$~K for Ga$_{0.91}$Mn$_{0.09}$N
\cite{Sono01}, is consistent with the expectations of the
mean-field Zener model \cite{Diet01d}. In more general terms,
within this model, large magnitudes of $T_{\mbox{\small {C}}}$
result from the combine effect of the large on-site p-d exchange
interaction and the efficient transfer of spin information owing to
a relatively large extend of the p-wave functions.

Another way pursed recently consists of searching for the high
temperature ferromagnetism in compounds containing combination of
elements from both II-VI and III-V materials. Layered
Ga$_{1-x}$Mn$_x$Se \cite{Peka98} and particularly chalcopyrite
Cd$_{1-x}$Mn$_x$GeP$_2$ \cite{Medv00} are starting to provide
encouraging results. Since, in general, III-V compounds can easier
be doped by impurities that are electrically active, whereas II-VI
materials support greater concentration of transition metals, a
suggestion has been put forward to grow magnetic III-V/II-VI short
period superlattice \cite{Kama01}, in which a charge transfer to
the magnetic layers will increase $T_{\mbox {\small C}}$.

\subsection{Beyond Mn-based compounds}

The highest spin and the associated large magnitude of the on-site
correlation energy $U$ account for the divalent character of the Mn
atoms in a large variety of environments. This results, in
particular, in a large solubility of Mn in II-VI materials and its
acceptor character in III-V compounds. A question arises about
ferromagnetic properties of semiconductors containing other
magnetic components. Quite generally, carrier-mediated exchange
mechanism is efficient if both concentration of holes and p-d
exchange energy $\beta N_0$ are appropriately large. According to
Fig.~1, for many transition metals and hosts, the $d$ states lie in
the gap. Hence, to trigger carrier-mediated exchange interactions,
co-doping by shallow impurities is necessary. However, shallow
acceptors will produce band holes only if the d$^{N+1}$/d$^N$ donor
level resides at lower energies, that is in the valence band.
Furthermore, except for some resonant situations, the magnitudes of
$|\beta N_0|$ are small for $N < 5$ because of opposite signs of
two relevant contributions \cite{Kacm01}. With these arguments in
mind, (Ga,Fe)N and (Ga,Co)N co-doped with acceptor impurities
appear as promising materials \cite{Blin02,Blin01}.

There are, however, other possible roads to high temperature
ferromagnetism. Since for configurations other than d$^5$, the
magnitude of $U$ is relatively small, the Hubbard-Mott transition,
and the associated ferromagnetism may appear at increasing density
of the magnetic constituent. To our knowledge no such effects have
so far been detected in materials in question. Furthermore, for
some magnetic ions and hosts, either d$^{N}$/d$^{N+1}$ or
d$^{N}$/d$^{N-1}$ level lies below the expected energy of shallow
donor or above energy of shallow acceptor, respectively. In such a
situation, the co-doping by shallow impurities will rather affect
the occupancy of the d levels than introduce carriers into the
bands, a known semi-insulating character of semiconductors
containing specific transition metals. At appropriately large
concentrations of both magnetic ions and additional impurities, the
carriers trapped on d levels may start to transfer, {\it via}
Zener's double exchange mechanism \cite{Zene51b}, magnetic
information between localised spins. Indications of ferromagnetism
in the vicinity of room temperature discovered in n-(Zn,Co)O
\cite{Ueda01b} and n-(Zn,V)O \cite{Saek01} may represent, according
to Fig.~1, such a case. Actually, a simultaneous doping by two
different magnetic impurities may serve to control both spin and
charge states as well as the carrier concentration in the specific
bands. The room-temperature ferromagnetism detected after doping a
few percent of Co to nonmagnetic TiO$_2$ \cite{Mats01a} may belong
to this category. In particular, the hybridized d levels of Ti and
Co may constitute the efficient channel for transmission of
magnetic information between the Co spins.

Finally, one should recall the existence of, e.g., ferromagnetic
europium chalcogenides and chromium spinels. In those compounds,
ferromagnetism is not mediated by carrier transport. With no doubt,
the availability of intrinsic and n-type tetrahedrally-coordinated
ferromagnetic compounds would enlarge considerably the impact of
semiconductor electronics. Actually, a theoretical suggestion has
been made \cite{Blin96a,Blin96b} that superexchange in Cr-based and
V-based II-VI compounds can lead to a ferromagnetic order. Desired
material properties, such as divergent magnetic susceptibility and
spontaneous magnetization, can also be achieved in the case of a
strong antiferromagnetic super-exchange interaction.  The idea here
\cite{Diet94b} is to synthesize a ferrimagnetic system that would
consist of antiferromagnetically coupled alternating layers
containing different magnetic cations, e.g., Mn and Co.

It is also tempting to consider the performance of DMS containing
ions from other groups of magnetic elements. Current works on
Si$_{1-x}$Ce$_x$ \cite{Fuji99} and on the insulator-to-metal
transition in amorphous Si$_{1-x}$Gd$_x$ \cite{Xion99} alloys can
be quoted in this context. However, because of a weak hybridization
between 4f and band states (exploited in Er-doped emitters), no
high-temperature ferromagnetism is expected in rare-earth-based
systems. More promising in this respect are materials containing 4d
or 5f ions --- there exists already a preliminary report on
spin-spin interactions in undoped Pb$_{1-x}$U$_x$Te \cite{Isbe95}.
Actually, in view of indications of room-temperature weak
ferromagnetism in (Ca,La)B$_6$ \cite{Youn99} and polimerized
rh-C$_{60}$ \cite{Maka01}, which are built up of non-magnetic
constituents, the spectrum of possibilities appears as unlimited.

The above list of achievements and prospects clearly shows that
searches for high temperature ferromagnetic semiconductors have
evolved into a broad field of materials science.  In addition to
work on design and synthesis of new systems, a considerable effort
will certainly be devoted to control contributions from
ferromagnetic or ferrimagnetic precipitates and inclusions in the
materials available already now. On theoretical side, the interplay
between Anderson-Mott localisation, disordered Stoner magnetism,
and carrier-mediated ferromagnetism will attract a considerable
attention. With no doubt we will witness many unforeseen
developments in the field of ferromagnetic semiconductors in the
years to come.

\section*{Acknowledgments}

The author is grateful to F. Matsukura and H. Ohno in Sendai and to
J. K\"onig and A.H. MacDonald in Austin for collaboration on III-V
magnetic semiconductors as well as to J.~Cibert, D.~Ferrand, and
S.~Tatarenko in Grenoble and to J.~Jaroszy\'nski, P.~Kossacki, and
M.~Sawicki in Warsaw for collaboration on II-VI magnetic
semiconductors. The work was supported by Foundation for Polish
Science, by State Committee for Scientific Research, Grant
No.~2-P03B-02417 as well as by FENIKS project (EC:
G5RD-CT-2001-00535).

\end{document}